# Accumulation-mode two-dimensional field-effect transistor: Operation mechanism and thickness scaling rule

*Nan Fang\* and Kosuke Nagashio\*\**
Department of Materials Engineering, The University of Tokyo, Tokyo 113-8656, Japan

ABSTRACT: Understanding the operation mode of a two-dimensional (2D) material-based field-effect transistor (FET) is one of the most essential issues in the study of electronics and physics. The existing Schottky barrier-FET model for devices with global back gate and metallic contacts overemphasizes the metal-2D contact effect, and the widely observed residual conductance cannot be explained by this model. Here, an accumulation-mode FET model, which directly reveals 2D channel transport properties, is developed based on a partial top-gate $MoS_2$ FET with metallic contacts and a channel thickness of 0.65~118 nm. The operation mechanism of an accumulation-mode FET is validated and clarified by carefully performed capacitance measurements. A depletion capacitance-quantum capacitance transition is observed. After the analysis of the $MoS_2$ accumulation-mode FET, we have confirmed that most 2D-FETs show accumulation-mode behavior. The universal thickness scaling rule of 2D-FETs is then proposed, which provides guidance for future research on 2D materials.

## INTRODUCTION

An accurate understanding of the operation mechanism of electronic devices is critical, especially for new channel materials, because the extraction of physical properties and the further control of the device characteristics are based on the operation mode. In recent years, transition metal dichalcogenide (TMD) field effect transistors (FETs) have attracted significant attention due to their potential application in ultimate scaled devices.[1-5] Typical TMD-FET devices are composed of a metallic source and drain contacts, and the metal/channel interfaces are under gate control, that is, a typical global back gate structure, as shown in **Figure 1a**. One of the key performance-limiting factors in 2-dimensional (2D) FETs is the 2D/metal contact.[6] Based on this idea and on the historical background of similar structures for carbon nanotubes,[7] ultrathin silicon on insulator (SOI),[8] and silicon nanowire FETs,[9] the Schottky barrier FET (SB-FET) model is proposed and developed to explain the 2D-FET operation mechanism.[10-12] Since the tunneling transport at the SB junction is dominant, studies on achieving low contact resistance by choosing metal types and inserting van der Waals materials and so on[13-15] are promoted. The most important success in SB-FETs is the explanation of the ambipolar behavior. However, this model oversimplifies the channel effect in many cases. Although the injected carriers from contact will inevitably be scattered through the commonly used micro-long channel, the scattering issues are often neglected in the SB-FET model. Moreover, the residual conductance observed in most multilayer 2D-FETs when over the critical thickness[16-31] cannot be explained by only the SB-FET model, suggesting a 2D depletion nature.[18,28] These contradictions suggest the existence of an additional operation mode focusing on the channel properties.

Here, to clarify this channel depletion-related operation mode, we focus on partial top-gate FETs with ohmic metallic contacts, where the 2D/metal contact is not modulated as shown in **Figure 1b**.[4,32,33] This type of device structure is often explained by accumulation mode (ACCU) FETs in a Si nanowire,[34] as shown in **Figure 1a**, where the gate controls the on and off states via accumulation and depletion of the majority of carriers in the partial gate region. The unipolar behavior is achieved due to *p/n* junction formation. However, the accumulation mode mechanism in 2D-FETs and Si nanowires has not been systematically investigated.

The fundamental technique to directly detect carrier density and interface states in semiconductors is capacitance measurements (*C-V*),[35-37] which provides critical insights to elucidate the 2D-FET operation mechanism. Although *C-V* measurements are quite informative, blindly applying this method established based on a bulk metal-oxide-semiconductor (MOS) FET/capacitor to an ACCU-FET can lead to incorrect conclusions. Experimentally, $C$-$V$ measurements in small-area 2D materials are very sensitive and always suffer from several difficulties.[5,38-41] A systematic study on the parasitic capacitance resulting from an $n^+$-Si/SiO$_2$ substrate and the channel resistance effect in $C$-$V$ is necessary. Theoretically, quantum capacitance in monolayer MoS$_2$ has been clarified in our previous work.[33] However, the study on capacitance transition from monolayer to bulk MoS$_2$ is still lacking.

In this work, mechanically exfoliated MoS$_2$ with a thickness from 0.65 (monolayer) ~ 118 nm is selected as the channel material for top-gate FET devices. Systematic investigations of $C$-$V$ and $I$-$V$ measurements are carried out for the same samples. For $C$-$V$, the parasitic capacitance is totally suppressed by using a quartz substrate. Frequency dispersion for low-mobility thin 2D channels mainly comes from the channel resistance effect. A transition from quantum capacitance ($C_Q$) to depletion capacitance ($C_D$) is observed from monolayer to bulk MoS$_2$. Having clarified the electrostatic field-effect control mechanism of carriers by $C$-$V$, the electrical transport data are explained by ACCU-FET for all channel thicknesses. The thickness scaling rule is proposed based on the ACCU-FET mechanism, which provide the complete picture of the transport properties for most of the 2D materials.

**RESULTS AND DISCUSSION**

*I-V characterization; increase in on-state conductivity and residual conductance.* **Figure 1a, b** shows a schematic drawing and optical image of the Al$_2$O$_3$ top-gate MoS$_2$ FET on the insulating quartz substrate. **Figure 2a** shows the typical conductivity ($\sigma$) – top gate voltage ($V_{TG}$) characteristic at $V_{DS}$ = 0.1 V with a MoS$_2$ thickness ($t_{MoS2}$) of 0.65, 16, 44, and 58 nm. It should be noted that $\sigma$ is normalized by the width and length without the thickness of MoS$_2$ flakes. Monolayer MoS$_2$ shows a clear off and subthreshold region. There are two distinct features observed by increasing the MoS$_2$ thickness. One is the increase in the on-state conductivity for the 16-nm-thick sample, which gradually saturates for thicker MoS$_2$ samples. The other is the abrupt increase in the residual conductance for the 44-nm-thick sample.

To focus on these two features, the maximum conductivity and the ratio of on-state to off-state current ($I_{ON}/I_{OFF}$) are shown in the range of $t_{MoS2}$ = 0.65 ~ 118 nm in **Figure 2b**. The maximum conductivity is controlled by the conductivity of the accumulation layer, not by the parasitic resistance since σ gradually increases with increasing $V_{TG}$, even for the 58-nm-thick sample. Therefore, it is discussed similarly to the mobility analysis in MoS$_2$. Instead of intrinsic photon scattering, coulomb scattering due to interfacial impurities is found to be dominant in the scattering mechanism of ultrathin MoS$_2$.[42-44] The extrinsic Debye length ($L_D$) is given here for the screening length of Coulomb scattering since most of the 2D materials are intrinsically charged by defects and impurities.

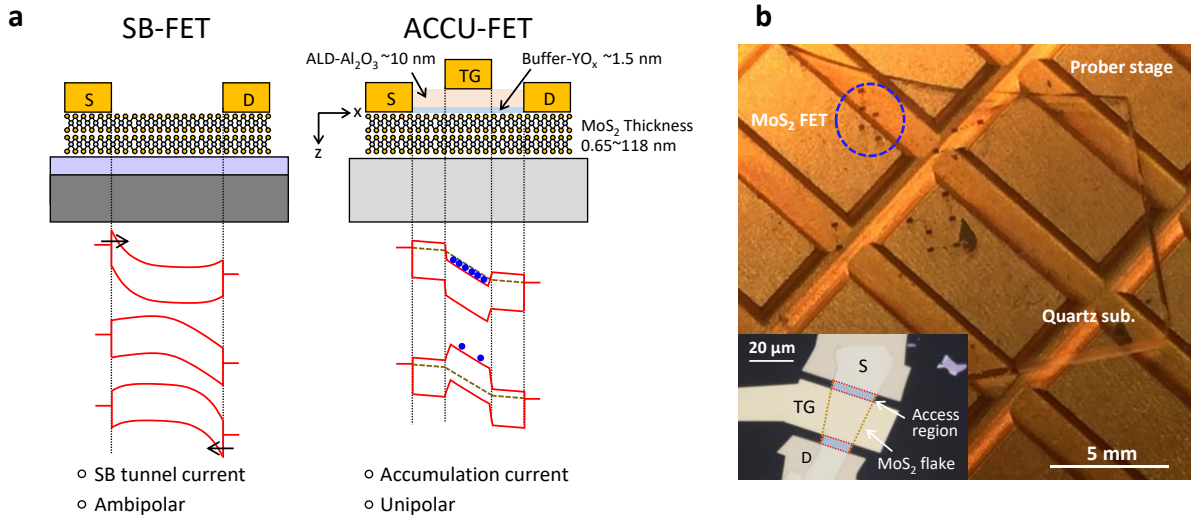

**Figure 1.** (a) Schematic of a back-gate SB-FET (left). Band diagrams of the *n*-branch, off state, and *p*-branch are shown below. Schematic of a top-gate ACCU-FET (right). Band diagrams of the accumulation and depletion states are shown below. The x and z directions in the Cartesian coordinate are defined. (b) Optical image of the device on a quartz substrate. The access region refers to the channel region uncovered by the top gate electrode. The existence of access region guarantees that contact region is not modulated by top-gate bias, and the effective channel length is defined by top-gate electrode width. S, D and TG indicate the source, drain and top gate electrodes, respectively.

$$L_D = \sqrt{\frac{\varepsilon_{MoS2}k_BT}{e^2 N_D}}. \quad (1)$$

$\varepsilon_{MoS2}$, $k_B$, $T$, and $e$ are defined as the dielectric constant of MoS$_2$ in the direction normal to the basal plane, the Boltzmann constant, the temperature, and the elementary charge, respectively. $N_D$ is the density of the donors (density of acceptors $N_A$ for p-type 2D). 2×$L_D$ is used in the following discussion to account for both the top and bottom interfaces. The MoS$_2$ with $t_{MoS2} > 2L_D$ will be undisturbed by the interfaces and maximum conductivity is saturated. Consideration of quantum-mechanical effect of accumulation capacitance ($C_A$) would give a more accurate carrier distribution in thick MoS$_2$ flakes.[45]

For $I_{ON}/I_{OFF}$, two regions are clearly observed: $I_{ON}/I_{OFF} > 10^5$ for $t_{MoS2}$ = ~0.65 - 35 nm and $I_{ON}/I_{OFF} < 10$ for $t_{MoS2} > 60$ nm. The transition occurs at $t_{MoS2}$ = ~48 - 55 nm. For ACCU-FET,[46] the conduction comes from "body current flow", which is modulated by the depletion region in the channel. The screening length ($\lambda_{ACCU\text{-}FET}$) is determined by the maximum depletion width ($W_{Dm}$), which can be expressed as follows:

$$\lambda_{ACCU-FET} = W_{Dm} = \sqrt{\frac{4\varepsilon_{MoS2}k_BT\ln(N_D/n_i)}{e^2 N_D}}, \quad (2)$$

where $n_i$ is intrinsic carrier density. $\lambda_{ACCU\text{-}FET}$ is independent of oxide capacitance ($C_{ox}$). An increase in the residual conductance occurs (e.g., 44-nm-thick sample in **Figure 2a**) when $t_{MoS2}$ becomes close to $W_{Dm}$ due to screening of the gate control. The present data indicates that $W_{Dm}$ is ~48 - 55 nm. It should be noted that this $W_{Dm}$ is roughly consistent with that in the previous data for global back gate MoS$_2$ FETs.[18,19] Generally, the global back gate 2D layered channel FET has been considered an SB-FET. In the case of SB-FETs, the off-state is achieved by controlling the barrier height at the MoS$_2$/metal contact independent of the channel thickness. This behavior is inconsistent with the SB-FET model. Moreover, for the global back gate 2D devices, the degradation of subthreshold swing (S.S.) with increasing channel thickness has been claimed as evidence for SB-FETs.[12] However, the similar degradation of S.S. is clearly observed due to the reduction in gate control by $t_{MoS2} \sim W_{Dm}$, as shown in

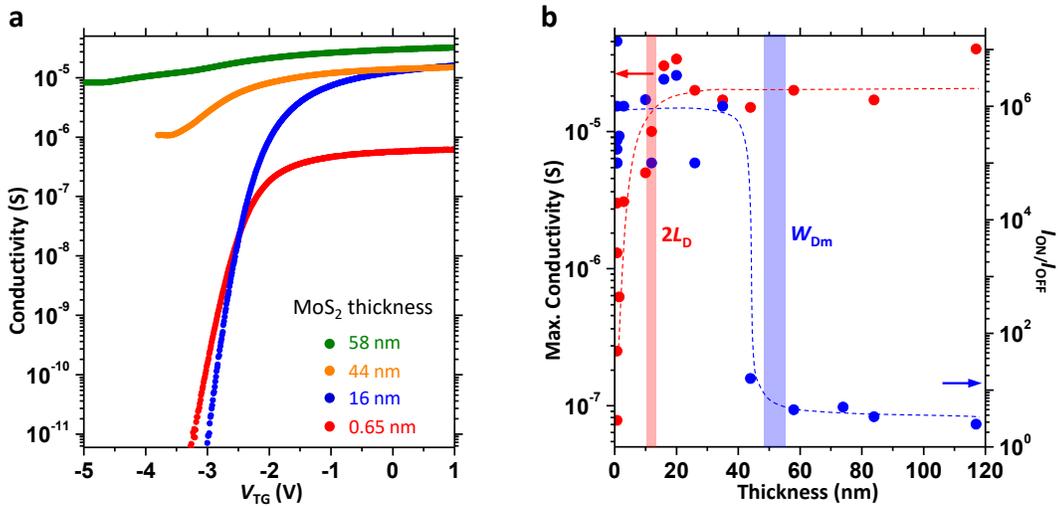

**Figure 2.** (a) σ–$V_{TG}$ characteristics at $V_{DS}$ = 0.1 V with a MoS$_2$ thickness of 0.65, 16, 44, and 58 nm. (b) Maximum conductance and $I_{ON}/I_{OFF}$ ratio as a function of thickness.

**Figure 2a**. These discussions suggest that ACCU-FET mode could exist in conjunction with SB-FET mode, even in widely fabricated back-gate 2D-FETs. In the following discussion, we use "bulk" for MoS$_2$ with $t_{MoS2} > W_{Dm}$ and "multilayer" for $t_{MoS2} < W_{Dm}$ for simplicity.

**C-V characterization; $C_Q$ & $C_D$.** To gain further insight into the operation mechanism of ACCU-MoS$_2$ FETs, C-V measurements with a frequency range of 1 k ~ 1 MHz are also conducted for the same devices. It should be noted that MoS$_2$ flakes with a large area (> 30 μm$^2$) were selected to improve the signal-to-noise ratio in the capacitance measurement. The full equivalent circuit used to model the top gate MoS$_2$-FETs is shown in **Figure 3a**. Here, $C_{para1}$ and $C_{para2}$ are the two types of commonly observed parasitic capacitance. $R_{access}$ is defined as the sum of MoS$_2$/metal contact resistance and MoS$_2$ resistance at the access region indicated in **Figure 1b**, which is constant. $R_{channal}$ is the MoS$_2$ channel resistance just below the top gate electrode and is modulated by the top gate bias. $C_{it}$ and $R_{it}$ are the interface states' capacitance and resistance, respectively, which account for carrier capture and emission processes. $C_{D(A)}$ and $C_Q$ are the focus of this paper, and they determine the carrier density in the conduction band (CB). $R_D$ is the resistance that models the supply of carriers to the depletion layer when $C_{D(A)}$ dominates the capacitance.

Several pitfalls are first discussed for MoS$_2$-FET-based C-V. The first pitfall is the parasitic capacitance effect, which comes from the widely used $n^+$-Si/SiO$_2$ substrate. As indicated previously,[47] in the double-gated geometry, there is capacitive coupling between back and top gates through the large contact pad area, which induces large parasitic capacitance. $C_{para1}$ refers to the parasitic capacitance that is charged or discharged through constant $R_{access}$. This will induce large frequency dispersion (>$C_{ox}$) in C-V and corresponding peaks in the conductance-frequency ($G_p/\omega$-$f$) measurement (Supplementary **Figure S1**). $C_{para2}$ refers to the parasitic capacitance that could shift the baseline of the C-V curve. A quartz substrate is used in this paper to totally remove these parasitic capacitances (Supplementary **Figure S2**). In this situation, $C_{ox}$ can be obtained in the strong

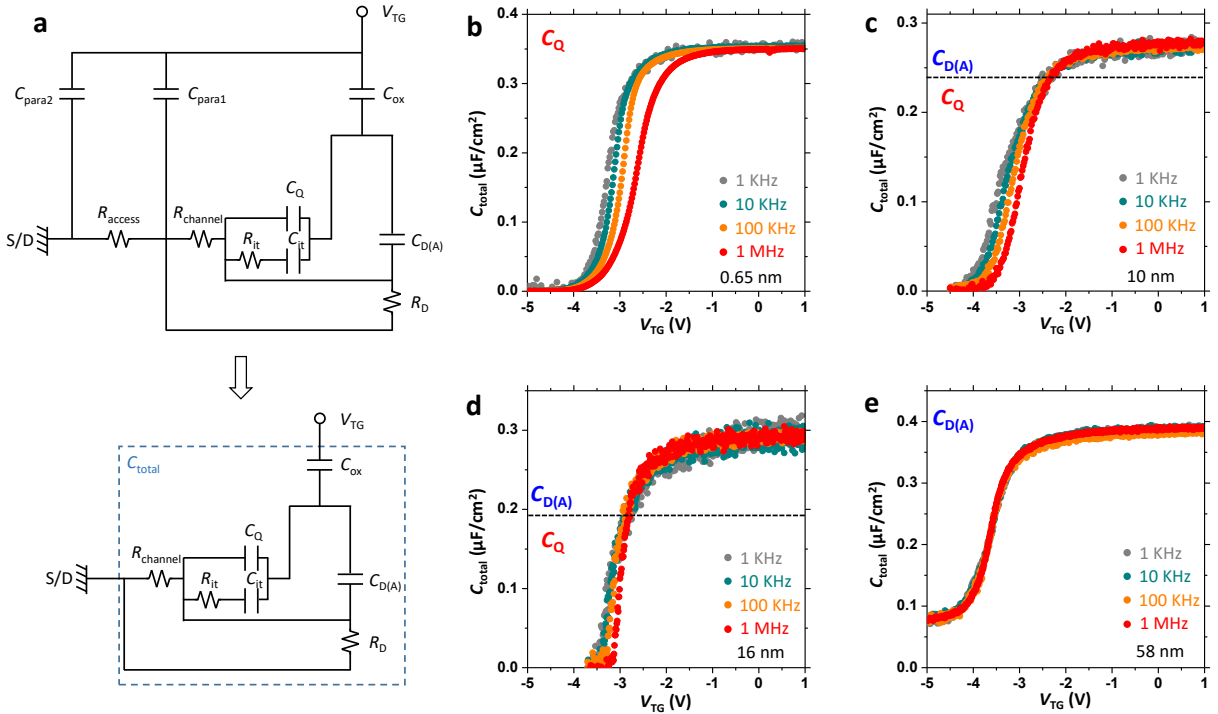

**Figure 3.** (a) Full and simplified equivalent circuits to model MoS$_2$-FET C-V. Strictly speaking, $C_Q$ and $C_D$ are not in parallel, because both come from majority carrier response in MoS$_2$. However, $C_Q$ and $C_{D(A)}$ are shown in parallel here in order to show charging path transition clearly. That is, if one exists, the other does not. (b-e) $C_{total}$-$V_{TG}$ characteristics of MoS$_2$ FETs with a thickness of 0.65, 10, 16, and 58 nm, respectively. Frequency ranges from 1 kHz to 1 MHz.

accumulation region when no frequency dispersion is observed.

The second pitfall is the access resistance effect, which could induce frequency dispersion in the accumulation region in C-V. $R_{access}$ is experimentally measured as the residual impedance at the high-frequency limit in the strong accumulation region where the other resistance is shunted. The measured $R_{access}$ is on the order of ~10 kΩ in most of the samples due to the natural n-doped property of MoS$_2$ and the low contact resistance with Ni. As shown in Supplementary **Figure S3** and **Note S1**, the $R_{access}$ effect in our measured frequency range can be neglected since $R_{access}$ is smaller than ~5×10$^4$ Ω. We have to mention that $R_{access}$ can still severely affect capacitance measurements at low temperature and for other 2D materials with higher resistance. Now, the equivalent circuit can be simplified (**Figure 3a**), where the experimentally measured source/drain to gate capacitance is defined as $C_{total}$.

**Figure 3b-e** shows the experimental $C_{total}$-$V_{TG}$ curves in the frequency range of 1 k - 1 MHz with $t_{MoS2}$ = 0.65, 10, 16 and 58 nm, respectively. Since the parasitic capacitance is totally removed by using the quartz substrate, the minimum capacitance plateau observed for $t_{MoS2}$ = 58 nm results from the contribution of $C_D$ with $W_{Dm}$. That is, the inversion layer is formed, resulting in a constant depletion width. The electrical communication still passes through free electrons at the edge of the depletion region because the p-n junction is formed between the inversion layer and ungated n-channel region. This C-V curve is consistent with that of a 1-μm-thick MoS$_2$ capacitor,[38] which also supports that $W_{Dm}$ is shorter than $t_{MoS2}$ = 58 nm. This cannot occur in SB-FET but is unique to the depletion behavior in ACCU-FET. As a result, the undepleted MoS$_2$ layer will always remain, which results in residual conductance and low $I_{ON}/I_{OFF}$ in I-V. On the other hand, for monolayer MoS$_2$, $C_Q$ contributes to $C_{total}$, instead of $C_D$. It originates from the partially occupied density of states (DOS) of CB modulated by the Fermi energy ($E_F$) in the Fermi-Dirac distribution.[48,49] Distinct from $C_D$, one of the main behaviors of $C_Q$ is that it follows an exponential decrease with respect to $E_F$ when $E_F$ is modulated in the band-gap. Due to the large band-gap of MoS$_2$, $C_Q$ can reach a small value, which results in an extremely low carrier density. This will be experimentally observed as a decrease to almost zero in C-V (**Figure 3b**) and a clear subthreshold/off region in I-V (**Figure 2a**). Although **Figure 3c, d** shows the transition from $C_Q$ to $C_D$, it is somewhat complicated. Therefore, it will be discussed later with the help of the quantitative analysis.

**Frequency dispersion by channel charging effect in C-V.** Before considering the transition from $C_Q$ to $C_D$ with increasing $t_{MoS2}$, $R_{channel}$ is discussed since it could induce frequency dispersion in the depletion region in C-V. Shockley-Read-Hall (SRH) theory is the basis to study carrier capture and emission process by the traps.[50] Based on this theory, a series $R_{it}$-$C_{it}$ network is modeled in the equivalent circuit, and experimental impedance spectroscopy always tries to capture this $R_{it}$-$C_{it}$-induced signal by excluding other capacitance or resistance effects. Large frequency dispersion is widely observed in the capacitance measurement of thin MoS$_2$ and other 2D-FET.[5,33,39,41] It is often treated as $R_{it}$-$C_{it}$-induced signals. However, other resistance effects could also introduce frequency-dependent signals. $R_{channel}$ is always parasitic in the FET structure, which cannot be avoided. In this section, $R_{channel}$ effect will be studied quantitatively. Monolayer MoS$_2$ is selected here because it shows the largest frequency dispersion and the simplest $C_Q$ expression.

**Figure 4a** shows the equivalent circuit of monolayer MoS$_2$ FET. $C_1$ is defined as the ideal capacitance by neglecting any resistance effect, and $C_1 = \frac{(C_Q+C_{it})C_{ox}}{C_Q+C_{it}+C_{ox}}$. Experimentally, when resistance exists in the equivalent circuit, it will give the R-C circuit, in which the time constant (τ) is determined. $C_{total}$ will decay from $C_1$ for $\omega\tau > 1$, where $\omega$ is angular frequency. $\tau_{Rch}$ and $\tau_{it}$ are defined as the time constants from $R_{channel}$ and $R_{it}$, respectively. **Figure 4b** shows measured $C_{total}$ as a function of frequency (C-f) at different $V_{TG}$ for the monolayer device in **Figure 3b**. The clear decay of $C_{total}$ at a specific frequency indicates that the capacitance is limited by one type of resistance.

For $\tau_{Rch}$, it is derived from a transmission line model[51,52] as follows (Supplementary **Figure S4, Note S2**):

$$\tau_{Rch} = \frac{C_1 R_{S,channel} L^2}{4}, \qquad (3)$$

where L is the channel length and $R_{S,channel}$ is the sheet resistance of MoS$_2$ channel. The drift current model[49]

is applied to express $R_{S,channel}$. Because the channel is on the order of micrometers in length and the drain bias is small, the diffusion current is negligible. Moreover, the drift current model reveals free carrier transport in the conduction band, which enables us to correlate C-V with I-V in the next part. $R_{S,channel} = \frac{1}{en_{ch}\mu}$, where $n_{ch}$ is the channel carrier density and $\mu$ is the drift mobility. $C_{it}$ and $\mu$ are extracted from the I-V characteristics[33] (Supplementary **Figure S5**). A higher $C_{it}$ means that more states need to be charged, which results in a larger $\tau_{Rch}$. On the other hand, $\tau_{it}$ is calculated based on SRH theory[53] in a 2-dimensional system as follows:

$$\tau_{it} = \frac{1}{\sigma_{capture-2D} v_{th} n_{ch}}, \quad (4)$$

where $\sigma_{capture-2D}$ is the capture cross section of interface states, which largely depends on the type of interface states. For point defects (*e.g.*, sulfur vacancy), it would be close to the atom size of ~0.3 nm. For band tail interface states induced by bond bending of Mo-*d* orbitals,[33] it could be on the order of 10 nm.[54] Therefore, $\sigma_{capture-2D}$ is assumed to be in the range of 0.3~10 nm. $v_{th}$ is thermal velocity of ~1.2×10$^7$ cm/s at room temperature by considering the electron effective mass of monolayer MoS$_2$ as $m^* = 0.6\ m_0$, where $m_0$ is the electron mass in a vacuum.

The calculated time constant as a function of $E_F$ is shown in **Figure 4c**. $\tau_{Rch}$ w/ $C_{it}$ is ~3 orders of magnitude larger than $\tau_{it}$ and is distributed across the measured frequency range of 1 kHz to 1 MHz, which indicates that the time constant due to the channel charging effect is the origin of the frequency-dependent capacitance behavior in **Figure 4b**. It is noted that both $\tau_{it}$ and $\tau_{Rch}$ with $C_{it}$ have a similar exponential $E_F$ dependence because the parameter $n_{ch}$ is included.

The experimental C-V and C-f curves are then reproduced by considering $R_{channel}$ instead of $R_{it}$. $C_{total}$ is derived as (Supplementary **Note S2**):

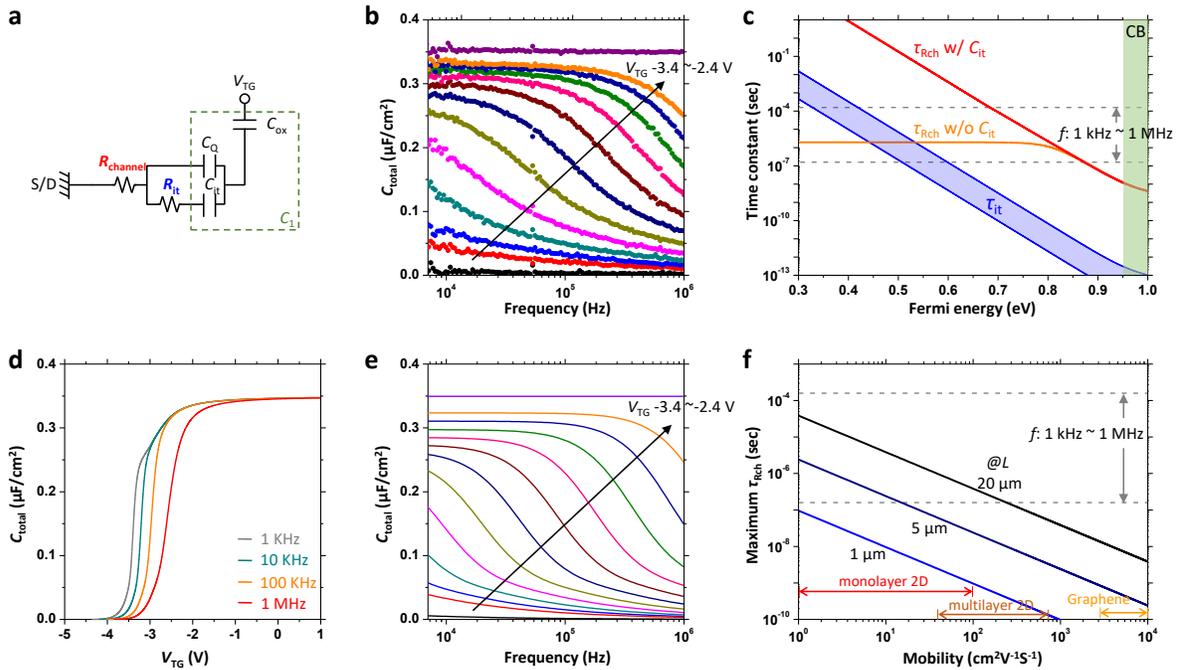

**Figure 4.** (a) Simplified lumped equivalent circuit to model monolayer MoS$_2$-FET C-V. Both $R_{channel}$ and $R_{it}$ could limit the frequency response of the capacitance. (b) Experimental $C_{total}$ as a function of frequency at different $V_{TG}$ (-3.4~-2.4 V) from monolayer MoS$_2$ FETs. A large frequency response is observed, which corresponds to the frequency dispersion of C-V in **Figure 3b**. (c) Calculated time constant as a function of $E_F$. Monolayer MoS$_2$ is assumed to have a bandgap of 1.9 eV. $E_F$ = 0.95 eV indicates the bottom of the CB. Dash line indicates the experimental measured frequency range by using $\tau=1/\omega$. (d) Calculated plot of $C_{total}$-$V_{TG}$ curves. (e) Calculated plot of $C_{total}$-$f$ curves at different $V_{TG}$ values (-3.4~-2.4 V). (f) Maximum $\tau_{Rch}$ as a function of mobility at different $L$.

$$C_{total} = C_1 \text{Re}\left[\frac{\tanh \lambda}{\lambda}\right],$$

where $\lambda = \sqrt{j\omega\tau_{Rch}}$. (5)

$V_{TG}$ is calculated as follows:

$$V_{TG} = V_{TG,mid-gap} + \int_0^{E_F/e}(C_{Q(D)} + C_{it} + C_{ox})/C_{ox}\,d(E_F/e).$$ (6)

$V_{TG,mid-gap}$ is a fitting parameter to compensate the MoS$_2$ $n$-doping effect. Equation (6) will be used to correlate $E_F$ with $V_{TG}$. Later, we will study multilayer MoS$_2$, where the surface potential $\psi_S$ is used instead of $E_F/e$. $C_{it}$ is included in equation (6) since the interface states always respond to the direct current (dc) $V_{TG}$. The simulation reproduces the experimental data quite well (**Figure 4 d,e**), suggesting that the experimentally observed frequency dispersion in $C$-$V$ does not result from the electron capture/emission process at the interface traps but from the channel charging effect. From the above study of $R_{channel}$ effect, let us review our previous work on the $C$-$V$ study of monolayer MoS$_2$.[33] $C_Q$ with a clear temperature dependence is correctly extracted since it is obtained at the strong accumulation region where $R_{channel}$ is shunted. Although the band-tail type energy distribution for the interface states is also reserved qualitatively, the widely used high-low frequency method on 2D-FET-based $C$-$V$ [5,33,39,41] will not reveal the true $C_{it}$ value quantitatively because the extracted time constant is indeed $\tau_{Rch}$ instead of $\tau_{it}$.

To provide guidance on how to avoid the channel charging effect in all 2D-FET-based $C$-$V$ with different thicknesses from monolayer to multilayer, the universal expression is derived. The region where $C_Q \ll C_{ox}$ should be considered since the $R_{channel}$ effect is severe due to the low carrier density. We assume that $C_{it}$ is smaller than $C_Q$, that is, attention should always be paid to improve the interface. In this case, $C_1 = C_Q$. Then, based on equations (3, 8) and the definition of $C_Q = \frac{dn_{ch}}{d\psi_S}$, $\tau_{Rch}$ will have a constant maximum, which is similar to $\tau_{Rch}$ without $C_{it}$ in the monolayer case (**Figure 4c**). This is because $n_{ch}$ in both $C_Q$ and $R_{S,channel}$ cancel with each other. This constant maximum is shown as follows:

$$\text{Maximum } \tau_{Rch} = \frac{L^2}{4\mu(k_BT/e)}.$$ (7)

The maximum $\tau_{Rch}$ is shown as a function of $\mu$ for various $L$ in **Figure 4f**. $\tau_{Rch}$ should be smaller than the measured frequency range to avoid the channel charging effect. For $L = 1$ μm, the allowable $\mu$ can be as low as 1 cm$^2$V$^{-1}$s$^{-1}$. However, due to both experimental difficulty and small signal-to-noise ratio, $L$ is usually in the range of 5 ~ 20 μm in our samples. In this case, $\mu$ is very important. $\mu$ is usually low in monolayer 2D materials, i.e., < 100 cm$^2$V$^{-1}$s$^{-1}$, at room temperature, while multilayer 2D materials have a higher $\mu$, which has the potential to avoid the channel charging effect. This has been confirmed in our 16-nm-thick device with suppressed frequency dispersion in **Figure 3d**. On the other hand, for graphene-based FETs, this effect can usually be neglected due to the extremely high $\mu$, which accounts for the recently observed frequency dispersion-free $C_{it}$ in a bilayer graphene/$h$-BN/graphite heterostructure.[55]

**$C_D$-$C_Q$ transition and MoS$_2$ ACCU-FET operation mechanism.** Now let us consider the $C_D$-$C_Q$ transition. Most of the measured MoS$_2$ FETs with $t_{MoS2} > 55$ nm show depletion capacitance (accumulation capacitance)-dominant $C$-$V$ without thickness dependence. $C_D$ and $C_A$ are separated by flat-band voltage ($V_{FB}$). Moreover, channel resistance-induced frequency dispersion is totally suppressed because $R_{channel}$ is shunted by the unmodulated conductive MoS$_2$ region, which results in a low charging resistance $R_D$. Thus, the equivalent circuit can be simplified as a lumped circuit, and conventional $C_{D(A)}$ analysis method can be applied. This enables us to extract parameters such as $N_D$ and $\varepsilon_{MoS2}$ of bulk MoS$_2$. The minimum $C_D$ is given as $\text{Minimum } C_D = \frac{\varepsilon_{MoS2}}{W_{Dm}}$. By considering that $W_{Dm}$ is 48~55 nm and the minimum $C_D$ is ~ 0.1 μF/cm$^2$, bulk $\varepsilon_{MoS2}$ is extracted as 6.3. This is roughly consistent with the calculated bulk $\varepsilon_{MoS2}$ in the z direction.[56] Based on equation (2), $N_D$ is determined to be 2~3×10$^{17}$ cm$^{-3}$. With these parameters, by using conventional $C_D$ expression (Equation (11) in Supplementary **Note S3**) and equation (6) without $C_{it}$, the $C$-$V$ of 58-nm-thick MoS$_2$ is fitted (**Figure 5a**). The simulated $C$-$V$ fits well with the experimental data. The slight deviation is due to $C_{it}$-induced distortion and the stretch-out effect.

$C_D$-$C_Q$ transition always occurs in multilayer MoS$_2$ FET-based *C-V*. Firstly, free electrons at the edge of the depletion region still communicate electrically with S/D through the ungated *n*-channel region. By modulating $V_{TG}$ negatively, the depletion width will reach $t_{MoS2}$ (16 nm). As a result, the electrical communication in *C-V* occurs between S/D and the quite small density of free electrons in the "depletion region". Based on this scenario, when the depletion width reaches $t_{MoS2}$, it can be considered that the $C_D$-$C_Q$ transition occurs, since the carrier density in the "depletion region" can be controlled by $C_Q$. Therefore, the *C-V* curve goes to zero even for the multilayer. After the whole channel is depleted, the surface potential will be continuously increased by further decreasing $V_{TG}$. Finally, the inversion layer will be formed. However, inversion capacitance corresponding with the *p*-branch in *I-V* cannot be observed because of the *p-n* junction, as schematically illustrated in **Figure 5b**.

Now, let us reproduce the *C-V* curve for $t_{MoS2}$ = 16 nm by simple analytical calculation. Since the expression for $C_{D(A)}$ is already obtained, the surface potential ($\psi_s$) is calculated in order to obtain the expression for $C_Q$ of multilayer MoS$_2$. The boundary condition of electric field = 0 at z = $t_{MoS2}$ is used for the solution of the one-dimensional Poisson equation. This is the intrinsic condition for the present 2D-FET structure, where the channel is always surrounded by the insulator or other insulating environment. The calculated potential distribution is shown in the inset in **Figure 5b**. By modulating the surface potential with the change of $\Delta\psi_s$, the potential in the channel changes everywhere ($\Delta\psi_z$) with the same value, that is, $\Delta\psi_z = \Delta\psi_s$, indicating that the whole channel can be fully controlled by $\psi_s$ and $\psi_s$ has a similar function as $E_F$ in monolayer $C_Q$ to modulate $n_{ch}$. With the calculated potential distribution, $C_Q$ is shown below (Supplementary **Note S3**):

$$C_Q = N_Q \exp\left(\frac{e\psi_s}{k_B T}\right), \quad (8)$$

where $N_Q$ is a constant independent of $\psi_s$. It is not surprising to see that $C_Q$ in the multilayer has a similar formula as that in the monolayer case with the same exponential $e/k_B T$ dependence.[49] Then, using $C_Q$ and $C_{D(A)}$ without $C_{it}$, the experimental data are well fitted, as shown in **Figure 5b**. The cross point indicates the transition from $C_D$ to $C_Q$ at $t_{MoS2} = W_D$. The slight deviation from the analytical $C_Q$ comes from the contribution of $C_{it}$. In **Figure 3b-e**, the transition from $C_D$ to $C_Q$ is clearly seen with decreasing MoS$_2$ thickness. Moreover, it is interesting that the frequency dispersion is observed only in the $C_Q$-dominant region. This is because the charging resistance $R_D$ is low enough for the $C_{D(A)}$-dominant region, while $R_{channel}$ is quite high for the $C_Q$-dominant region.

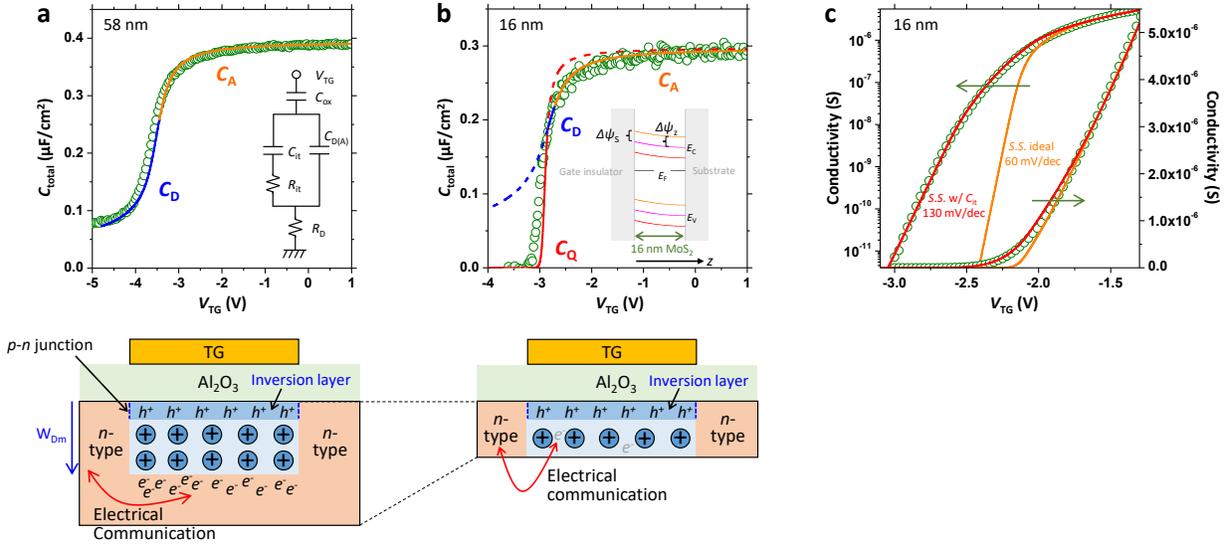

**Figure 5.** (a) Depletion capacitance (accumulation capacitance) of a 58-nm-thick MoS$_2$ FET. The green circle is the experimental $C_{total}$-$V_{TG}$ curve at 1 MHz from **Figure 3e**. The blue and orange solid lines are the theoretical fitting curve based on $C_D$ and $C_A$ ($C_{total} = \frac{C_{D(A)}C_{ox}}{C_{D(A)}+C_{ox}}$), respectively. The inset is the simplified lumped equivalent circuit to model bulk MoS$_2$-FET *C-V*. The bottom schematic shows the situation for $V_{TG} \ll 0$ V. (b) $C_D$-$C_Q$ transition in a 16-nm-thick MoS$_2$ FET. The green circle is the experimental $C_{total}$-$V_{TG}$ curve at 1 MHz from **Figure 3d**. The blue, orange and red solid lines are the theoretical plots based on $C_D$, $C_A$ and $C_Q$ ($C_{total} = \frac{C_Q C_{ox}}{C_Q + C_{ox}}$), respectively. The inset shows the potential distribution calculated as a function of MoS$_2$ thickness (z direction) at different values of $\psi_s$. The bottom schematic shows the situation for $V_{TG} \ll 0$ V. (c) Transfer characteristics of a 16-nm-thick MoS$_2$ ACCU-FET. The green circuit is the experimental σ-$V_{TG}$ curve from **Figure 2a**. The orange and red solid lines are the ideal theoretical fitting curves without $C_{it}$ and with $C_{it}$, respectively.

A large advantage for *C-V* on the FET structure is that it directly estimates the carrier density in the transport phenomenon of *I-V*. Meanwhile, for *C-V* on the capacitance structure, the potential distribution in the channel is affected by the additional back metal contact, and the whole depletion channel cannot be obtained.[57] Having theoretically calculated all of the components in $C_{total}$, it is possible to further reproduce the *I-V* characteristics by introducing the drift current model. $n_{ch}$ is calculated as $n_{ch} = \int C_{Q/D(A)} d\psi_S$. The $V_{TG}$-$\psi_s$ relation is again calculated from equation (6) (parameters are shown in Supplementary **Figure S5**). The simulation result of this 16-nm-thick MoS$_2$ sample is shown in **Figure 5c**. The ideal *S.S.* of ~60 mV/dec can be obtained without the $C_{it}$ effect, which comes from the thermal limitation in $C_Q$. The experimental *I-V* from the off state to the linear region is then fully reproduced by including the $C_{it}$ effect. *S.S.* is degraded to ~130 mV/dec as well as a gradual transition from linear to the subthreshold region in the linear region. In the ACCU-FET, the equation to describe *S.S.* is given equivalently as in the conventional MOSFET as:

$$S.S. = \ln 10 \frac{k_B T}{e} \frac{C_{ox} + C_{it}}{C_{ox}}. \qquad (9)$$

This equation is valid from monolayer to multilayer when $t_{MoS2} \ll W_{Dm}$. $C_Q$ does not appear in this equation since it is much smaller than $C_{it}$ and $C_{ox}$ at the *S.S.* region. *S.S.* will be degraded when $t_{MoS2}$ becomes close to or larger than $W_{Dm}$ due to losing the gate control of the whole channel, as shown in **Figure 2a**. Combined with the *I-V* analysis in monolayer MoS$_2$ (Supplementary **Figure S5**) as well as the *C-V* analysis, we have successfully clarified the operation mechanism of MoS$_2$ ACCU-FET from monolayer to bulk flakes.

**Thickness scaling rule.** Compared with SB-FETs, where the tunneling transport at the SB contact junction is dominant, the channel transport properties are straightforwardly revealed for the present ACCU-FETs. The carrier density modulation with the drift current model used above gives a complete picture of carrier band transport under the gate field-effect condition. Now, the transport mechanism is divided into three regions as a function of $t_{MoS2}$. When $t_{MoS2} > W_{Dm}$, the channel is only partially controlled by the gate and shows band transport. The existence of residual conductance is the sign of this region. When $2L_D < t_{MoS2} < W_{Dm}$, the channel is fully controlled with optimized mobility because of screening of interfacial Coulomb scattering. Band transport also dominates in this region. When $t_{MoS2} < 2L_D$, band transport is still dominant at room temperature, but it often suffers from mobility degradation due to prominent interfacial Coulomb scattering. In the subthreshold region at low temperature, the localized states induced transport such as hopping will become dominant.[58-60] It should be noted that both $W_{Dm}$ and $2L_D$ are independent of $C_{ox}$, which enables us to propose the thickness scaling rule of transport properties for various 2D materials as a function of $N_D$ ($N_A$) (**Figure 6**). The summarized 2D materials here have a band-gap of 1~2 eV and a similar dielectric constant. A WSe$_2$ global back-gate FET on a SiO$_2$ (90 nm)/$n^+$-Si substrate was fabricated and characterized for comparison (Supplementary **Figure S6**). Although the observation of both $n$- and $p$-branches is explained by the SB-FET model, the increase in the maximum conductance and residual conductance with increasing WSe$_2$ thickness also reveals ACCU-FET behavior. As mentioned before, transport properties for the present top gate MoS$_2$ FET are consistent with that from global back gate MoS$_2$ devices. Therefore, almost all of the data on $W_{Dm}$ and $L_D$ in **Figure 6** are obtained from global back gate devices in the previous literature. At high $N_D$ ($N_A$) region (>10$^{19}$ cm$^{-3}$), $W_{Dm}$ will decrease substantially, resulting in a small thickness window for "fully controlled band transport", that is, fully depleted. In fact, full control of channel will be lost when the 2D thickness become greater than $W_{Dm}$. Moreover, it will be more degraded by considering a heavy doping effect such as band gap narrowing.[61] This explains why well-controlled FETs with high $I_{ON}/I_{OFF}$ are difficult to achieve in recent heavily doped 2D materials such as PtS$_2$, PtSe$_2$, SnS, and SnSe. Meanwhile, $2L_D$ is scaled down to just several atomic layers of thickness. This strong electrostatic confinement effect combined with increased $N_D$ ($N_A$) will introduce strong scattering. Band transport is difficult to achieve in atomically thin flake of these heavily doped 2D materials, and the Anderson localization phenomenon is suggested to be observed.[62] Moreover, in terms of 2D/metal contact, heavily doped 2D materials generally show low contact resistance because of the thin Schottky barrier width. From the above analysis, well controlled doping approaches on 2D crystals are in great demand for improving the performance of 2D ACCU-FET.

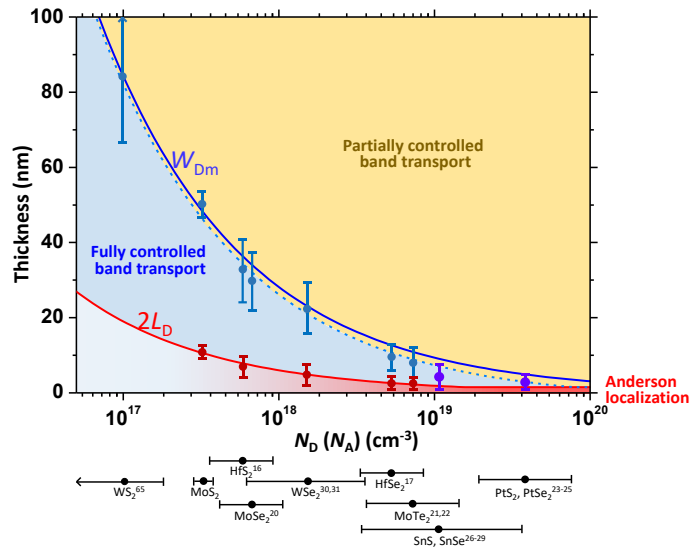

**Figure 6.** Thickness scaling rule of a 2D ACCU-FET. Baselines of $W_{Dm}$ and $2L_D$ are calculated from the parameters of bulk MoS$_2$. Band-gap = 1.29 eV and ε = 6.3. With a scaling thickness of 2D from bulk to monolayer, three regions will be observed. They are divided by $W_{Dm}$ and $2L_D$. Different types of 2D materials are shown here as a function of $N_D$ ($N_A$). Most of the 2D materials summarized here come from the mechanical exfoliation method, which gives a relatively stable $N_D$ ($N_A$). This situation might be different when using synthesis approaches.

**CONCLUSION**

As a conclusion, the top gate MoS$_2$-FETs are found to work at accumulation-mode, whose operation mechanism is clarified by capacitance measurement with special precautions. The ACCU-FET study here provides a new platform and analytical mode for the electronics and physics of novel nanomaterials. Moreover, the universal thickness scaling rule of 2D-FETs is proposed in terms of $W_{Dm}$ and $L_D$, which is applicable to most of semiconductor 2D materials.

**Methods**

MoS$_2$ films were mechanically exfoliated onto the insulating quartz substrate from natural bulk MoS$_2$ flakes. Ni/Au was deposited as source/drain electrodes. Then, Y metal with a thickness of 1 nm was deposited via thermal evaporation of the Y metal in a PBN crucible in an Ar atmosphere with a partial pressure of 10$^{-1}$ Pa, followed by oxidization in the laboratory atmosphere to form the buffer layer.[63, 64] The Al$_2$O$_3$ oxide layer with a thickness of 10 nm was deposited via atomic layer deposition, followed by the Al top-gate electrode formation. Raman spectroscopy and atomic force microscopy (AFM) were employed for determining the flake thickness. $I$–$V$ and $C$–$V$ measurements were conducted using Keysight B1500 and 4980A LCR meters, respectively. All electrical measurements were performed in a vacuum prober with a cryogenic system.

**ACKNOWLEDGEMENT**

N. F. was supported by a Grant-in-Aid for JSPS Research Fellows from the JSPS KAKENHI. This research was supported by the JSPS A3 Foresight program, Core-to-Core Program, A. Advanced Research Networks, JSPS KAKENHI Grant Number, JP16H04343, Japan.

**SUPPORTING INFORMATION**
Details of $C_{para1}$ effect on capacitance measurement; Details of $C_{para2}$ effect on capacitance measurement; Details of analysis on $R_{access}$ effect in capacitance measurement; Details of transmission line model in MoS$_2$ FET; Used parameters in the calculation; Transfer characteristics of back-gate WSe$_2$ FET; Details of analysis on channel resistance effect in capacitance measurement; Detailed analyses of $C_D$-$C_Q$ transition. This material is available free of charge *via* the Internet at http://pubs.acs.org.

## AUTHOR INFORMATION
**Corresponding Author**
Email: * nan@adam.t.u-tokyo.ac.jp, ** nagashio@material.t.u-tokyo.ac.jp

**Notes**
The authors declare no competing financial interests.

**Supporting Information**.

# Accumulation-mode two-dimensional field-effect transistor: Operation mechanism and thickness scaling rule

*Nan Fang*[*] *and Kosuke Nagashio*[**]

Department of Materials Engineering, The University of Tokyo, Tokyo 113-8656, Japan

[*] nan@adam.t.u-tokyo.ac.jp, [**] nagashio@material.t.u-tokyo.ac.jp

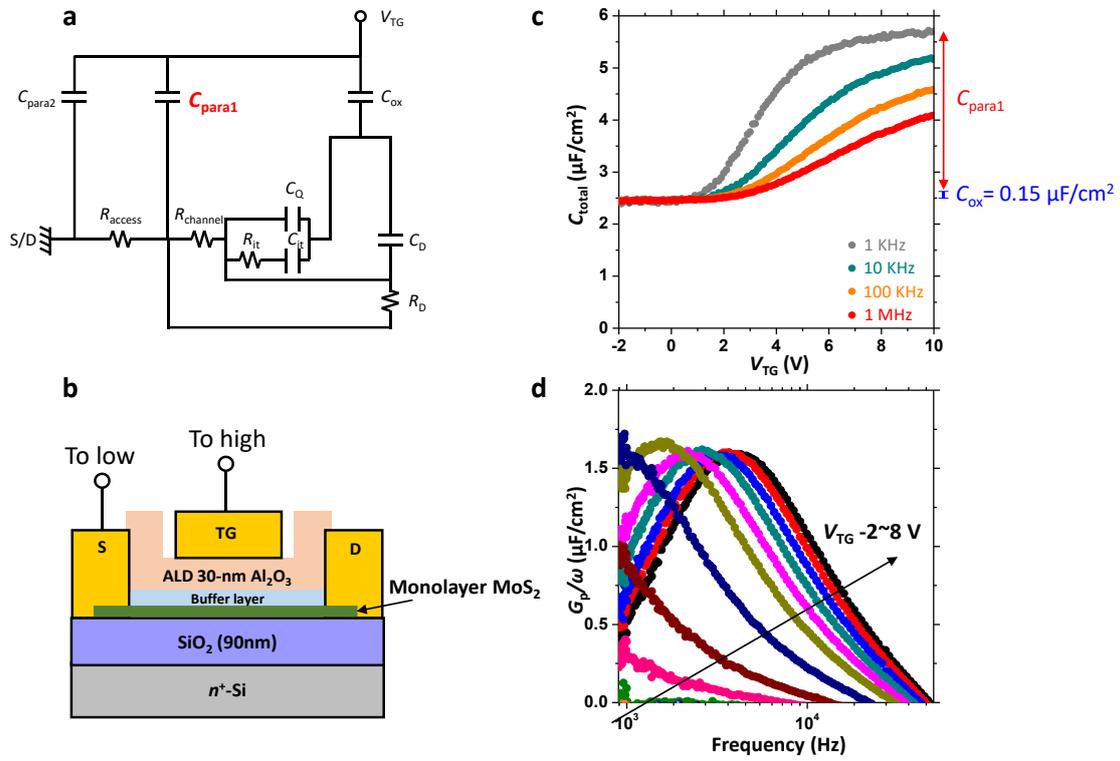

**Supplementary Figure S1.** (a) Full equivalent circuit to show $C_{para1}$. (b) The configuration of capacitance measurement. Source is connected to low-terminal and drain is floating. Top gate is connected to high-terminal. This configuration combined with $n^+$-Si/SiO$_2$ substrate introduce large frequency dispersion ($>C_{ox}$) in (c) C-V and (d) corresponding peaks in conductance-frequency ($G_p/\omega$-$f$) measurement. It should be noted that these peaks are not related with the interface traps.

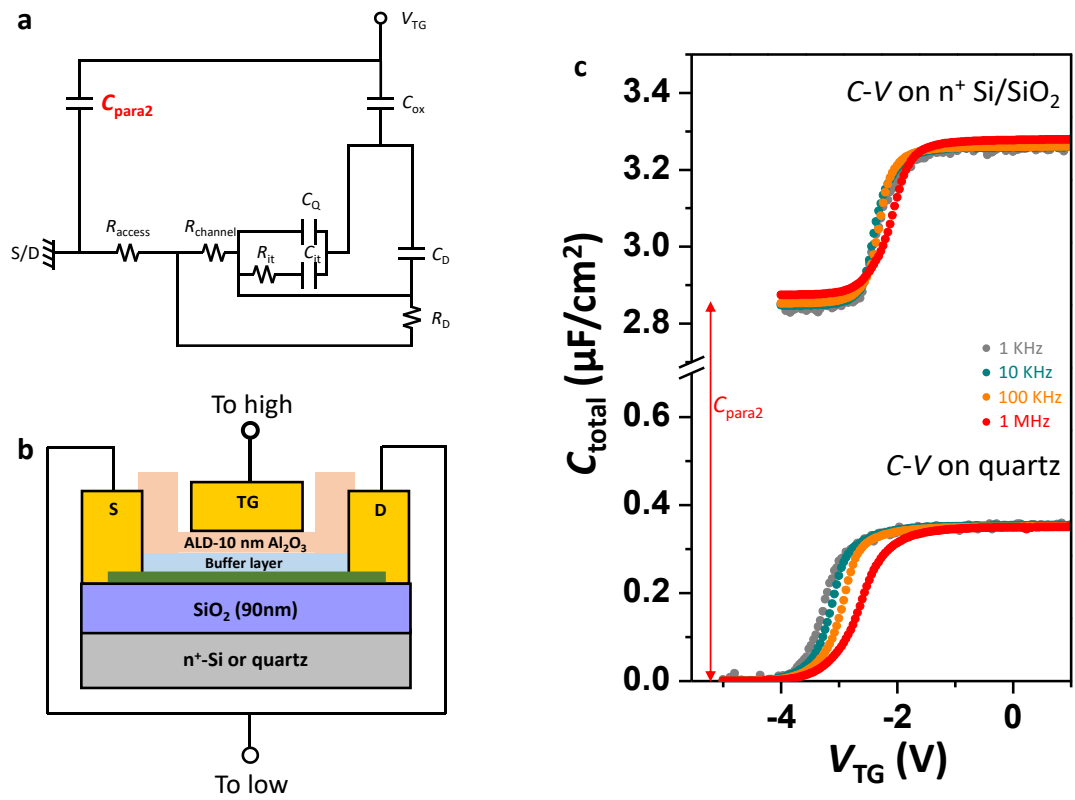

**Supplementary Figure S2.** (a) Full equivalent circuit to show $C_{para2}$. (b) The configuration of capacitance measurement. Both source and drain are connected to low-terminal and top gate is connected to high-terminal. Under this configuration, $C_{para1}$ can be suppressed even on $n^+$-Si/SiO$_2$ substrate. However, $C_{para2}$ still induces large baseline shift in C-V, which cannot be removed as shown in (c). By fabricating MoS$_2$ FET on quartz substrate, all the parasitic capacitance is totally removed.

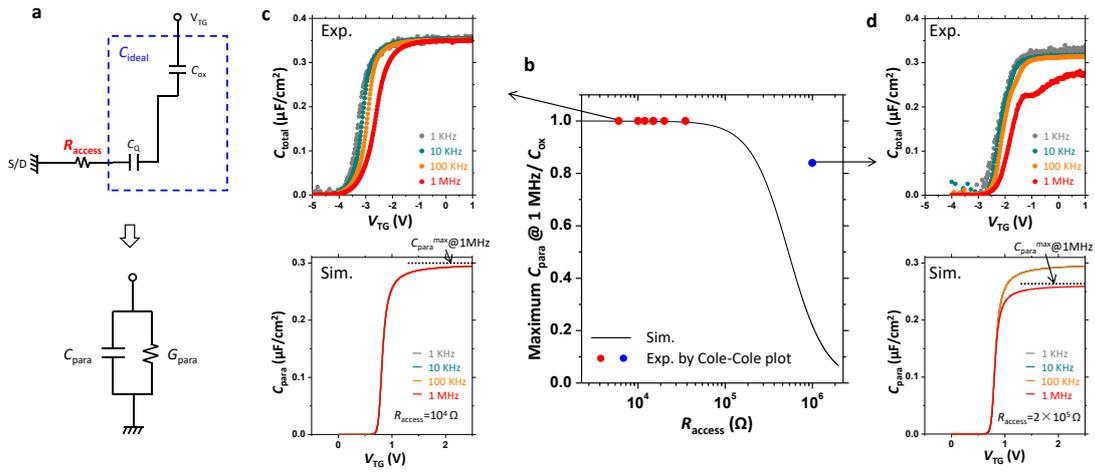

**Supplementary Figure S3.** (a) Simplified equivalent parallel circuit to study $R_{access}$ effect. (b) Maximum $C_{para}$ at 1 MHz/$C_{ox}$ as a function of $R_{access}$. (c,d) Experimental and simulated $C_{total}$-$V_{TG}$ characteristics of MoS$_2$ FET with different $R_{access}$. The decrease in Maximum $C_{para}$ at 1 MHz is clearly observed in (d) due to high $R_{access}$ ~1000 kΩ.

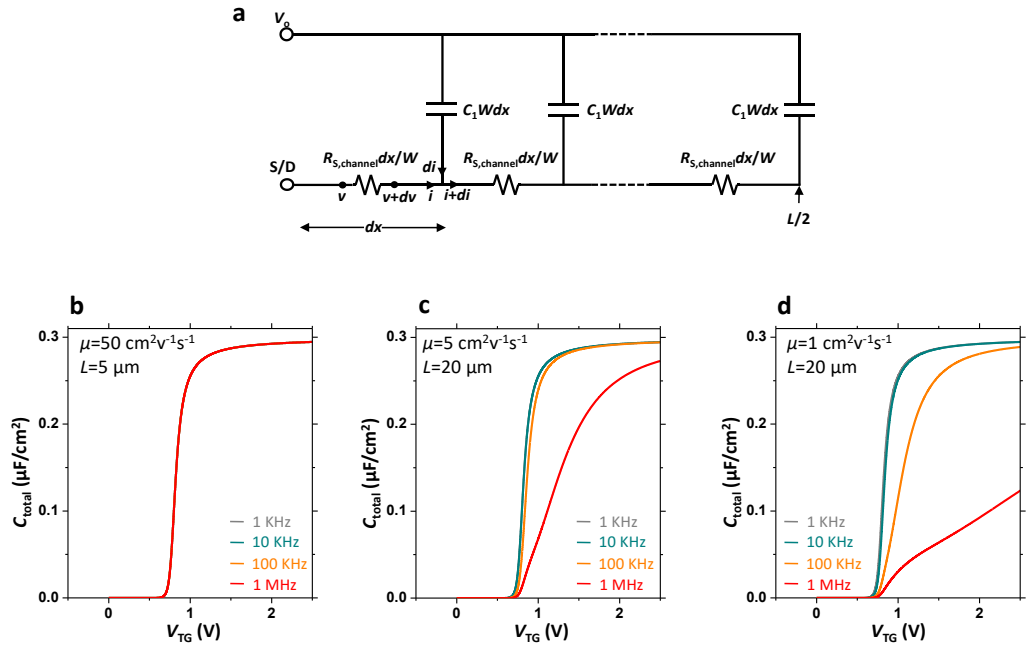

**Supplementary Figure S4.** (a) Transmission line equivalent circuit of MoS$_2$ FET. This equivalent circuit is valid when the whole channel of MoS$_2$ is depleted. (b,c,d) Calculated plot of $C_{total}$-$V_{TG}$ curves without $C_{it}$ at different $\mu$ and $L$. Although frequency dispersion is observed, experimental $C_{total}$-$V_{TG}$ curves in Figure 3b of the main text cannot be reproduced without band tail shape $C_{it}$.

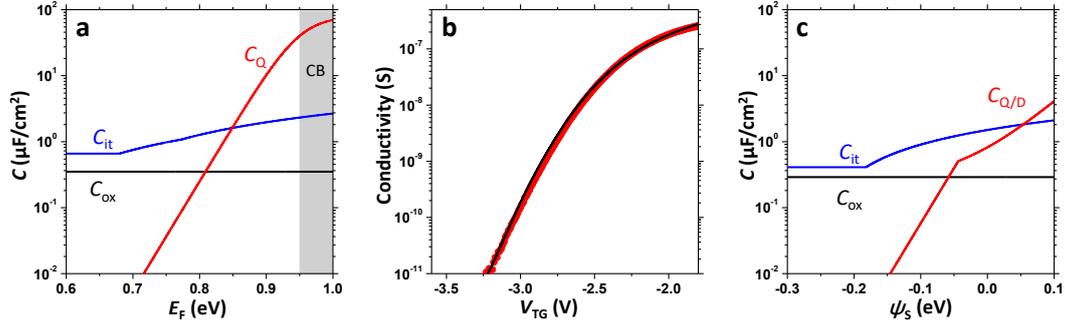

**Supplementary Figure S5.** (a) $C_Q$, $C_{it}$ and $C_{ox}$ used in monolayer MoS$_2$ *I-V* and *C-V* simulation. $C_{ox}$ is experimentally extracted from $C_{total}$ at strong accumulation region. $C_Q$ is theoretically calculated. $C_{it}$ and drift mobility are extracted from *I-V* fitting.[1] Drift mobility is slightly higher than conventional experimentally extracted field-effect mobility due to reduced carrier controllability of the gate by $C_{it}$. But field-effect mobility extraction is still one of the fastest way to study carrier transport properties. Drift mobility $\mu$ is assumed to be independent of $E_F$ with a constant value for simplicity. Here for studied monolayer MoS$_2$, $\mu = 2.2$ cm$^2$ V$^{-1}$ s$^{-1}$. This is a rough assumption because $\mu$ is usually dependent on carrier density through screening effect. But the dominant factor in determining $I_{DS}$ is the carrier density instead of the drift mobility especially at subthreshold region. This explains why we can give a good *I-V* fitting even at constant $\mu$ condition. (b) Experimental and calculated σ-$V_{TG}$ curve of monolayer MoS$_2$ FET. Red dot is experimental result and solid black line is simulated result by using parameters in (a). (c) $C_{Q/D}$, $C_{it}$ and $C_{ox}$ used in 16 nm-thick MoS$_2$ FET *I-V* and *C-V* simulation. Here for studied 16 nm-thick MoS$_2$, $\mu = 60$ cm$^2$ V$^{-1}$ s$^{-1}$.

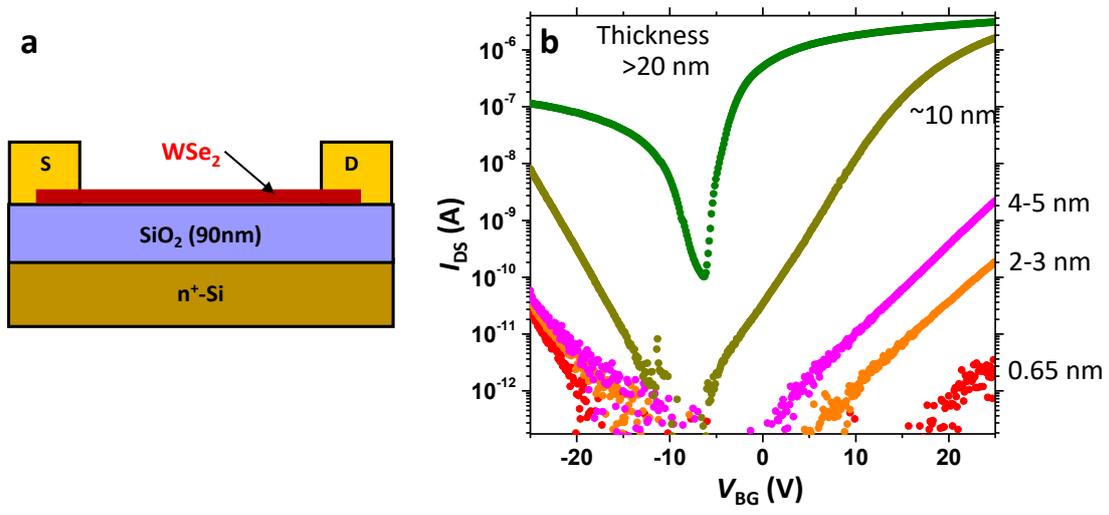

**Supplementary Figure S6.** (a) Schematic diagram of the device. (b) Transfer characteristics of back-gate WSe$_2$ FET with different channel thickness.

**Note S1. Access resistance effect on capacitance measurement.**

$R_{access}$ can cause error in measured capacitance. $R_{access}$ effect is simulated based on equivalent parallel circuit and other resistance effect is not included in this simulation as shown in Supplementary **Figure S3a**. Here, $C_{ideal}$ is defined as ideal C-V in monolayer MoS$_2$ without $C_{it}$. Change of $C_{Q/D}$ will not affect the conclusion in $R_{access}$ simulation. $C_{ox}$ is given as 0.3 μF/cm$^2$ with 10×10 μm$^2$ area. Equivalent parallel capacitance is shown as follows:

$$C_{Para} = C_{ideal} / (1 + \omega^2 C_{ideal}^2 R_{access}^2). \tag{1}$$

As we can see in Supplementary **Figure S3c,d**, the large error occurs in the accumulation region of C-V. Experimentally, access resistance can be extracted by Cole-Cole plot at accumulation region where other resistance is shunted. Maximum $C_{para}$ at 1 MHz remains to $C_{ox}$ when R$_{access}$ = 10$^4$ Ω and decrease when R$_{access}$ = 2×10$^5$ Ω. So maximum $C_{para}$ at 1 MHz/$C_{ox}$ is the good parameter to indicate $R_{access}$ effect. This parameter equals to 1 when $R_{access}$ can be neglected at measured frequency range and decrease when $R_{access}$ limit the measured capacitance as shown in Supplementary **Figure S3b**.

**Note S2. MoS$_2$ Channel resistance effect on capacitance measurement.**

Transmission line model has been applied to study channel resistance effect on C-V for Si MOSFET.[2,3] Here, transmission line model will also be used in MoS$_2$ FET to model $R_{channel}$ effect (Supplementary **Figure S4**). Notice that substrate is insulating in MoS$_2$ FET, which simplifies the mathematical expressions of equivalent circuit by neglecting charge supply from the substrate. Assume that all variables are in phasor quantities. $v_0$ refers to small ac variation. $R_{S,channel}$ refers to sheet resistance of R$_{channel}$. $i$ refers to current from one side (source or drain) of the electrode. So the total current from both sides $i_{D,S}$ is 2×$i$.

Firstly,

$$\frac{dv}{dx} = -\frac{R_{S,channel}}{W} i, \tag{2}$$

$$\frac{di}{dx} = j\omega C_1 W (v_0 - v). \tag{3}$$

Differentiating eq. 2 with respect to x and substituting it into eq. 3,

$$\frac{dv^2}{dx^2} = -j\omega C_1 R_{S,channel} (v_0 - v) \tag{4}$$

Assume $u = v - v_0$, we have:

$$\frac{du^2}{dx^2} = \gamma^2 u, \tag{5}$$

where $\gamma = \sqrt{R_{S,channel} j\omega C_1}$.

The solution of this eq. 5 is

$$u = Ae^{-\gamma x} + Be^{\gamma x} \tag{6}$$

Based on the boundary conditions,

$v = 0$ when $x = 0$ and $\frac{dv}{dx} = 0$ when $x = \frac{L}{2}$, we have:

$$A = \frac{v_0}{e^{-\gamma L}+1}, B = -\frac{e^{-\gamma L}v_0}{e^{-\gamma L}+1}. \tag{7}$$

Based on eq. 2 at $x = 0$, source/drain current:

$$i_{DS} = 2i = -\frac{2W}{R_{S,channel}}\frac{dv}{dx} = LWj\omega C_1 v_0 \frac{\tanh\frac{L\gamma}{2}}{\frac{L\gamma}{2}} \tag{8}$$

The propagation constant $\lambda$ and channel resistance limited time constant $\tau_{Rch}$ are given as below:

$$\lambda = \frac{L\gamma}{2} = \sqrt{j\omega\tau_{Rch}}, \tau_{Rch} = \frac{C_1 R_{S,channel} L^2}{4}. \tag{9}$$

So the experimental measured equivalent parallel capacitance and conductance $C_{total}$ and $G_{total}$ are:

$$C_{total} = C_1 \text{Re}\left[\frac{\tanh\lambda}{\lambda}\right], \frac{G_{total}}{\omega} = -C_1 \text{Im}\left[\frac{\tanh\lambda}{\lambda}\right] \tag{10}$$

**Note S3. $C_D$-$C_Q$ transition.**

Classical depletion capacitance (accumulation capacitance) expression of MoS$_2$ is given below by neglecting holes,[4]

$$C_{D(A)} = \frac{\varepsilon_{MoS2}}{\sqrt{2}L_D}\frac{\left|\exp(\frac{e\psi_S}{k_B T})-1\right|}{\sqrt{\exp(\frac{e\psi_S}{k_B T})-\frac{e\psi_S}{k_B T}-1}}. \tag{11}$$

Notice that often used $E_F$ in monolayer MoS$_2$ discussion is unsuitable in multilayer and bulk MoS$_2$ since potential $\psi$ changes from surface to body. Instead, surface potential $\psi_S$ is used.

By neglecting free carriers in the depletion region, the depletion layer width is shown as:

$$W_D = \sqrt{\frac{2\varepsilon_{MoS2}\psi_S}{eN_D}}. \tag{12}$$

As for bulk MoS$_2$ (Thickness > 55 nm), $W_{Dm}$ is obtained when $\psi_S$ saturates at strong inversion region. While for multilayer MoS$_2$ (Thickness < 35 nm), transition from depletion capacitance to quantum capacitance occurs when $W_D$ reaches MoS$_2$ body thickness. Transition condition of $\psi_S$ is given as below:

$$\psi_{S(D\to Q)} = \frac{eN_D t_{MoS2}^2}{2\varepsilon_{MoS2}}. \tag{13}$$

Thickness of MoS$_2$ is defined as $t_{MoS2}$. Quantum capacitance is calculated based on Poisson equation and boundary conditions.

The potential distribution is firstly calculated. Parabolic function is used due to Poisson equation as shown:

$$\psi_X = C_0 + C_1 z + C_2 z^2 . \tag{14}$$

One dimension Poisson equation is given as below:

$$\frac{d^2\psi_z}{dz} = \frac{-eN_D}{\varepsilon_{MoS2}} = A . \tag{15}$$

Two boundary conditions are given below:

(1) Potential at surface ($z=0$) is $\psi_S$.

$$\psi_z\big|_{z=0} = \psi_S . \tag{16}$$

(2) The electric field at $z=t_{MoS2}$ is approximately zero, which is due to insulating quartz substrate.[5]

$$\frac{d\psi_z}{dz}\bigg|_{z=t_{MoS2}} = 0 . \tag{17}$$

This condition is the intrinsic condition for present 2D-FET structure, where channel is always surrounded by the insulator or environment and is important for obtaining $C_Q$ dominant region. Either additional making metal contact to MoS2 such as capacitor structure or $t_{MoS2}>W_{Dm}$ will degrade this condition to conventionally used one, that is $\psi\infty=0$.

Using eq.14-17, potential distribution is obtained as

$$\psi_z = \frac{Az^2}{2} - At_{MoS2}z + \psi_S = f(z) + \psi_S . \tag{18}$$

Then free electron statistics is calculated. Free electron density per unit volume at position $z$ is written as

$$n(z) = N_D \exp(\frac{e\psi_z}{k_B T}) . \tag{19}$$

By integrating eq. 19 with z from 0 to $t_{MoS2}$, and replacing $\psi_z$ with eq. 18, we get total channel free electron per unit area as:

$$n_{ch} = N_D \exp(\frac{e\psi_S}{k_B T}) \int_0^{t_{MoS2}} \exp(\frac{ef(z)}{k_B T}) d_z . \tag{20}$$

Quantum capacitance is calculated from the definition $C_Q = \frac{dn_{ch}}{d\psi_s}$, which is given as below:

$$C_Q = \frac{en_{ch}}{k_B T} = N_Q \exp(\frac{e\psi_S}{k_B T}), \text{ where constant } N_Q = \frac{eN_D \int_0^{t_{MoS2}} \exp(\frac{ef(z)}{k_B T}) d_z}{k_B T} . \tag{21}$$

To simplify the calculation of $N_Q$, capacitance continuity condition is finally used. That is

$$C_D = C_Q \big|_{\psi_S = \psi_{S(D \to Q)}}. \tag{22}$$

Capacitance in multilayer MoS$_2$ ($C_{\text{MoS2}}$) is given by two piece-wise functions combined with eq. 11,13,21,22,

$$C_{MoS2} = \begin{cases} C_D, \psi_S > \psi_{S(D \to Q)} \\ C_Q, \psi_S < \psi_{S(D \to Q)} \end{cases}. \tag{23}$$